\documentclass[sigconf]{acmart}
\usepackage{hyperref}
\usepackage{subcaption}
\usepackage[utf8x]{inputenc}
\usepackage[main=english]{babel}
\usepackage{amsmath}
\usepackage{marginnote}
\usepackage{amssymb}
\usepackage{xcolor}
\usepackage{colortbl}
\usepackage{physics}
\usepackage{xspace}
\usepackage{flushend}
\usepackage{placeins}
\usepackage{tikz}
\usepackage{pgfplots, pgfplotstable}
\usetikzlibrary{shapes,backgrounds}
\usepackage{graphicx}
\usepackage{pdfpages}
\usepackage{soul}
\usepackage{float}

\newlength\figureheight
\newlength\figurewidth
\pgfplotsset{compat=1.14}

\newcommand{\nop}[1]{{}}

\newcommand{\eg}{\emph{e.g.}\xspace}

\newcommand{\QUBO}{\textsc{Qubo}\xspace}
\newcommand{\QUBOs}{\textsc{Qubo}s\xspace}
\newcommand{\FPTAS}{\textsc{Fptas}\xspace}
\newcommand{\PTAS}{\textsc{Ptas}\xspace}
\newcommand{\APX}{\textsc{Apx}\xspace}

\sethlcolor{black}

\begin{document}
\copyrightyear{2020}
\acmYear{2020}
\acmConference{\color{white}{}}
\acmBooktitle{}
\acmPrice{}
\acmDOI{}
\acmISBN{}

\title{Approximate Approximation on a Quantum Annealer}


\author{Irmi Sax}
\affiliation{%
  \institution{Technical University of Applied Science Regensburg}
  }
\email{irmengard.sax@othr.de}

\author{Sebastian Feld}
\affiliation{%
  \institution{LMU Munich, Mobile and Distributed Systems Group}
  }
\email{sebastian.feld@ifi.lmu.de}

\author{Sebastian Zielinski}
\affiliation{%
  \institution{LMU Munich, Mobile and Distributed Systems Group}
  }
\email{sebastain.zielinski@ifi.lmu.de}

\author{Thomas Gabor}
\affiliation{%
  \institution{LMU Munich, Mobile and Distributed Systems Group}
  }
\email{thomas.gabor@ifi.lmu.de}

\author{Claudia Linnhoff-Popien}
\affiliation{%
  \institution{LMU Munich, Mobile and Distributed Systems Group}
  }
\email{linnhoff@ifi.lmu.de}

\author{Wolfgang Mauerer}
\affiliation{%
  \institution{Technical University of Applied Science Regensburg}
  \institution{Siemens AG, Corporate Research}
  }
\email{wolfgang.mauerer@othr.de}
\renewcommand{\shortauthors}{I.~Sax et al.}

%
\begin{abstract}
Many problems of industrial interest are NP-complete, and quickly exhaust 
resources of computational devices with increasing input sizes. Quantum annealers (QA)
are physical devices that aim at this class of problems by exploiting quantum mechanical
properties of nature. However, they compete with efficient heuristics and probabilistic or
randomised algorithms on classical machines that  allow for finding approximate solutions
to large NP-complete problems.

While first implementations of QA have become commercially available, their \emph{practical}
benefits are far from fully explored. To the best of our knowledge, approximation
techniques have not yet received substantial attention.

In this paper, we explore how problems' approximate versions of varying degree can be
systematically constructed for quantum annealer programs, and how this influences
result quality or the handling of larger problem instances on given set
of qubits. We illustrate various approximation techniques on both, simulations
and real QA hardware, on different seminal problems, and interpret the results to
contribute towards a better understanding of the real-world power and limitations
of current-state and future quantum computing.


\end{abstract}

\maketitle

\section{Introduction}
%
Many industrial problems belong to complexity class NP, the set of problems solvable with
a non-deterministic Turing machine in polynomial time. Even if there are efficient algorithms
for small instances using fixed parameter or approximation algorithms~\cite{ren19,cyg19,dev19}
there are no known  algorithms to solve large enough instances exactly \emph{and} efficiently.

Quantum annealing is one candidate that might solve such hard problems in less time than
classical machines. The exact relations between classical and quantum 
computational power still pose many open questions, despite recent popular results that
prove quantum supremacy for certain very specific problems~\cite{aru19}. In particular,
it is not expected that quantum computers will be able to solve NP-complete problems
in polynomial time~\cite{aar13}. 

Industrial use-cases rarely focus on decision problems, but often concern approximate
optimisation. For instance, a practical use of the travelling salesperson problem (TSP)
is not to ask ``\emph{is there a closed route of length \(n\) between cities?}'' (which would
be captured by the NP-complete decision version of problem) , but rather
``\emph{what are possible short routes?}'', as described by the NPO version.
Accepting slight deviations from optimal solutions can lead to substantial
savings in temporal effort for many problems, which is usually preferable in 
pratical applications. 

First instances of commercially available QAs have appeared~\cite{mcg14,cas17,fel11};
expected theoretical and practical applications are numerous including quantum chemistry~\cite{str19}, traffic management~\cite{sto19}, network design~\cite{din19} or quantum assisted learning~\cite{wil19}. However, the precise benefits of QA caused by the
drastic shift in hardware implementation are mostly still unchartered territory~\cite{gyo19}.

\begin{figure}
\centering
  \includegraphics{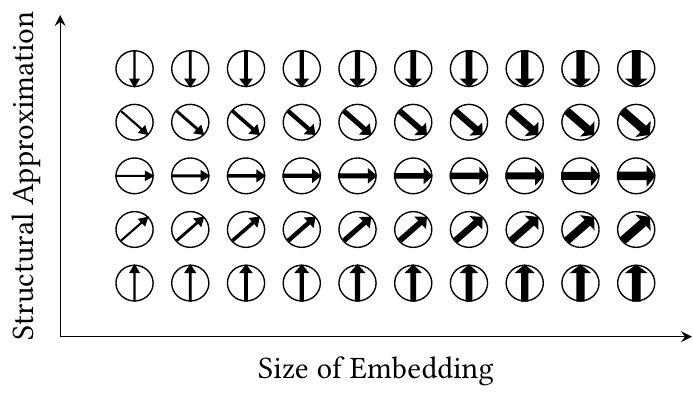}\vspace*{-1em}
  \caption{Illustration of solution quality dependent on the embedding's size (indicated by growing thickness of arrows) and degree of structural approximation (indicated by direction of arrow, whereas up represents good and down bad quality)}
\label{vergleich}
\end{figure}

In this paper, we discuss how programs for QAs can be designed to find non-optimal solutions to 
NPO problems, and what gains and losses in terms of various qualities are implied.
In classical computation, approximation algorithms usually sacrifice
precision for decreased runtime as compared to an exact
algorithm. Quantum annealers feature constant computation time by
nature of their design (see Section~\ref{cha:basics}),
and therefore need to trade other factors in relation to closeness to
optimality. There are two major opposing influences on solution
quality, as Figure~\ref{vergleich} illustrates. First, quality
depends how well logical qubits can be mapped to the available
physical qubits: Not all physical qubits are interconnected, and
interactions between unconnected qubits cause increased distance from
solution optimality.\footnote{If logical qubits do not match the structure of physical qubits they can be represented by chains, that is one logical qubit is mapped to several physical ones. If qubits of a chain have different binary values at the end of the annealing process the solution cannot be valid anymore. That is why we aim for short chains when mapping qubits to the hardware.~\cite{ven15}}  Second, exact mathematical formulations lead to
perfect solutions on flawless hardware, but complicate the mapping of
logical to physical qubits because exact specification usually requires
large amounts of qubits. Pruning the problem---for instance by
relaxing some of the constraints---is enabled by approximating the problem instance and thus approximating the solution but can also induce opposite effects on real hardware thanks to the smaller amount of required qubits.
One of the tasks we address in this work, is to find a
balance between an easy mapping of qubits to the hardware, and 
approximating problems appropriately such that the overall
probability of finding good solutions is amplified.

\section{Basics}
\label{cha:basics}

Quantum annealing differs considerably from classical notions of algorithmic
calculation. Therefore, we briefly summarise the core concepts in this section
(readers who wish to remind themselves of the detail can consult standard
references like, \eg~\cite{mcg14},~\cite{nie00}).

Quantum annealers solve minimisation problems specified as quadratic unconstrained
binary optimisation problems (\QUBO).
A \QUBO is a representation of a
minimisation formula using binary variables $x_i$. For a problem graph
$G_P = (V,E)$ with nodes $V = \{1,\ldots, n\}$ and weights $c_i$ for all
nodes and weights $c_{ij}$ for all edges in $E$, the \QUBO is defined
as~\cite{lew17}
\(\min \left(\sum_i^n c_ix_i + \sum_{(i,j) \in E} c_{ij}x_ix_j\right)\).
A \QUBO can equivalently be specified
by all nodes, edges and weights of a problem graph, or by the adjacency matrix of the problem graph. For an introduction on how to formulate a \QUBO for a given optimisation problem we refer the reader to~\cite{glo18}.

We perform experiment evaluations on the D-Wave \textit{DW2000 Q21}
quantum annealer~\cite{mcg14}. Conceptually, the device comprises two
components: A processor for solving mathematical problems by quantum
mechanical processes, and a user front-end that allows for controlling
the processor.

The chip provides about 2000 qubits in the structure of a
\emph{chimera} graph \(G_{C}\)~\cite{dwa18}. 16 $\times$ 16 cells are placed in a
quadratic grid, and each cell contains a graph with eight nodes. Every node
represents a qubit and is connected with four other qubits in the
same cell. Different cells are also interconnected, but not on the
level of each individual qubit.

To find solutions for a device-independent \QUBO $G_P$ on the 
hardware, $G_P$ must be mapped to $G_C$, referred to as \emph{embedding}.
The limited connectivity of $G_C$ with at most six edges per node implies that
a node in $G_P$ of higher degree cannot directly be embedded, and must be
represented by several nodes of $G_C$.
The amount of qubits requires in  \(G_{\text{C}}\) is therefore larger
than \(G_{\text{P}}\), which makes it desirable to formulate problems
such that the involved qubits require only low connectivity. For finding a proper embedding we used the tool \emph{minorminer} by D-Wave Systems Inc.\footnote{\url{https://github.com/dwavesystems/minorminer}}
		
		
		
		


\section{Approximating QUBOs}\label{cha:simplifying}
A fully specified QUBO delivers an exact, optimal solution of the encoded
problem if the underlying minimisation problem is solved. A QUBO matrix without
non-zero entries does not impose any constraints on the problem
variables, and every possible assignment represents a valid solution---in other words, an empty QUBO delivers a set of random binary values as result
of an optimisation process.

Interpolating between these scenarios intuitively gives solutions that
contain an increasing amount of random choices (and consequently,
deteriorating solution quality), while less qubits (non-zero
entries in the QUBO) are required to represent the problem.
Removing \QUBO entries that represent 
optimisation constraints leads to solutions that are usually not
optimal, but valid. Removing entries representing hard constraints often
results in invalid solutions. In this context, hard constraints describe all values in a \QUBO matrix that represent actual constraints on a solution. If those constraints are violated the solution is not valid and therefore not usable.  Consequently, we will focus on problems where
most of the \QUBO entries are optimisation constraints. Pruning a \QUBO induces 
two contrary effects (cf.~Figure~\ref{vergleich}): Removing entries 
implies a less exact specification, and \emph{decreases} solution 
quality, but sparsely populated \QUBO matrices are easier to embed on 
real QA hardware, which \emph{increases} solution quality.

We have devised different strategies that produce increasingly
approximate versions of a given QUBO:

\begin{description}
\item[\emph{Fraction}] This methods orders the non-diagonal entries by increasing
value. We delete blocks in entries in granularity of 5\% of all non-zero
values, up to 100\%, relative to all entries representing minimisation constraints. We delete small entries first because they are considered to have little effect on the optimal solution and will not approximate the problem instance significantly.
Entries representing hard constraints remain in the \QUBO, which also
holds for the other methods.
\item[\emph{Threshold}] This method determines the largest non-diagonal entry. The
smallest entry within 5\% difference from this value is taken as a threshold.
All values \emph{below} this threshold are pruned from the QUBO. Subsequent approximation
steps raise the threshold in increments of 5\%.
\item[\emph{Random}] This method starts with deleting 5\% of randomly chosen \QUBO entries
(entries representing hard constraints are always kept), and increases the amount of deleted
entries to 10\%, 15\%, etc. for increased degrees of approximation.
\end{description}

Note that eventually, all three methods arrive at identical QUBOs when 100\% of the
non-hard constraints are purged.

\begin{figure*}
		\centering 
		\includegraphics{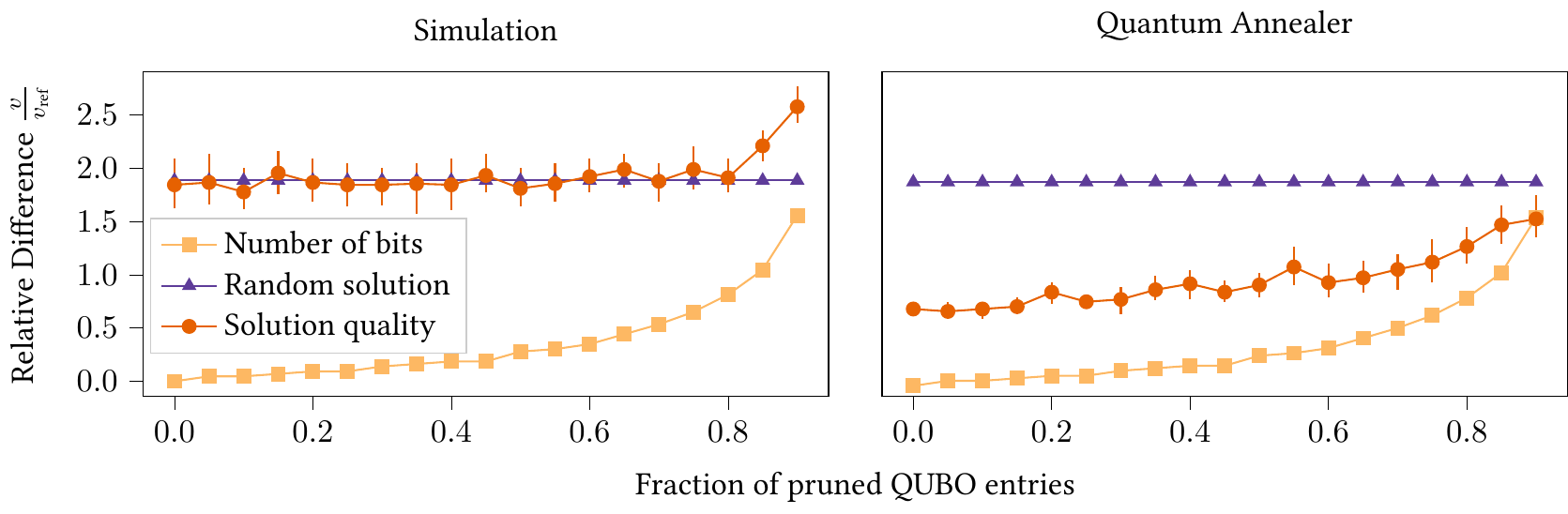}\vspace*{-1em}
        \caption{Ratio of errors ($v$) compared to number of elements in set $U$ ($v_{\textrm{ref}}$) and ratio of embed-able instance size ($v$) compared to the original, non-pruned instance ($v_{\textrm{ref}}$) for the Exact Cover Problem}
	\label{erg:ec}\vspace*{-1em}
\end{figure*}

We focus on removing entries representing optimisation constraints. If these
occur on non-diagonal entries in the \QUBO matrix, deleting leads to reduced
connectivity, shorter qubit chains and thus a smaller amount of required physical
qubits. However, deleting optimisation constraints from the diagonal of the \QUBO
matrix while leaving non-diagonal values in the corresponding matrix row or column
does not decrease connectivity, and hence does not reduce the amount of required
physical qubits. Therefore, we focus on problems where optimisation constraints 
reside in non-diagonal entries.
As all the problems analysed in this paper are computational problems from Karp's 21 NP-complete problems, the proposed approximation of those problems might not only be of industrial but also scientific interest.

\subsection{Exact Cover}
The \emph{Exact Cover} (EC) Problem considers a set $U$ of numbers and a set $V$ of subsets $V_i$. 
An exact cover for $U$ is a collection $V' \subset V$ of sets such that every element $u \in U$
appears exactly once in $V'$. The decision variant of Exact cover is in
NP~\cite{kar72}. One variation of the objective function is to minimize the number of errors:
An error occurs if an element of $U$ appears more than once in $V'$, or not at all.
Following Ref.~\cite{luc14}, the optimisation variant of the exact cover problem is given by
\begin{equation}
\min \sum_{u \in U} \left(1- \sum_{i: u \in V_i} x_i\right)^2 
\end{equation}
The binary variable $x_i$ is set $1$ if subset $V_i$ is contained in $V'$, and $0$ otherwise. A straight-forward
reduction from EC to Maximum-Weight Independent Set (MIS) is given by Choi~\cite{cho10}.

To discriminate hardware imperfections from consequences of the approximation proper, we run experiments
for each target problem on both, a classical simulation (the \emph{dimod.Simulated Annealing Sampler} by D-WAVE Systems Inc.\footnote{\url{https://docs.ocean.dwavesys.com/en/latest/docs\_dimod/reference/sampler\_composites/samplers.html\#module-dimod.reference.samplers.simulated\_annealing}}) and a quantum annealer. For the EC, our data set
requires about 1000 physical qubits to implement the exact \QUBO and was pruned step-wise to 53 qubits.
Only non-diagonal \QUBO entries have been deleted, which represents the information which elements of $U$ appear in which sets $V_i$. Every pruned \QUBO was solved 100 times. Results are shown in Figure~\ref{erg:ec}. \(v/v_{\text{ref}}\) is the 
relative difference between the number of errors \(v\) of the cover compared to a reference value $v_{\textrm{ref}}$, which is the size of set $U$.

\begin{figure*}
		\centering 
		\includegraphics[scale=1.0]{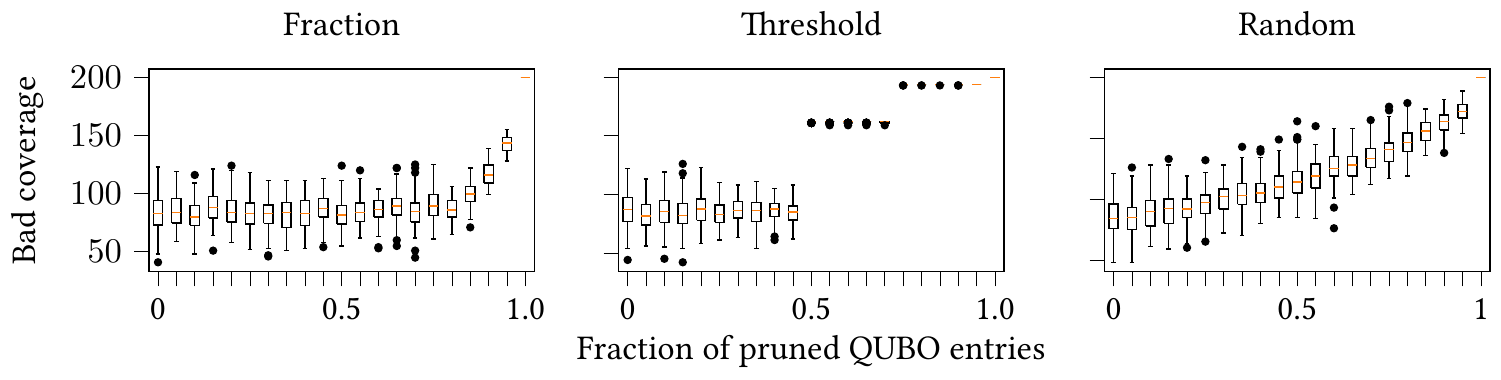}
        \caption{Results for 100 runs of the Exact cover problem on a simulation with different pruning methods}\label{ec:prun}\vspace*{-1em}
\end{figure*}
Pruned \QUBOs require less physical qubits, which in turn means that with increased
amounts of pruning, larger problem instances can be embedded on a given amount of
physical qubits. Figure~\ref{erg:ec} also shows the largest size of a pruned instance that can still be represented on the 2048-qubit QA hardware,
with the size -- in terms of logical qubits -- of the unpruned original instance. For the random solution we assigned the 53 \emph{logical} bits randomly  with values 0 and 1 for 100 times. The resulting mean error is also shown in Figure~\ref{erg:ec}. As a random solution is not influenced by the approximation of the problem instance but only the number of logical bits needed, it stays constant throughout the pruning process.

Deleting entries from the \QUBO matrix with method \emph{fraction} results in a constant error
for the simulation up to a pruning fraction of about 80\%. Removing more than 80\% of the \QUBO information leads to significantly worse solutions
The QA behaves similarly with increasing amount of pruning, save for a lower \emph{absolute} deviation value from the reference $v_{\textrm{ref}}$. The solution quality is mostly
unchanged up to a pruning factor of about 60\%, before it decreases significantly. In both
cases, the size of embeddable problems increases essentially linearly in the
regime of invariant solution quality.

Figure~\ref{ec:prun} compares the effect of different pruning strategies on the
solution (anti-)quality, measured in errors for a cover. The leftmost graph contains the same information as the left-hand part of Figure~\ref{erg:ec}, except that the distribution of
results for iterative runs is shown, compared to the mean value on Fig.~\ref{erg:ec}.

\begin{figure*}
    \centering
    \includegraphics{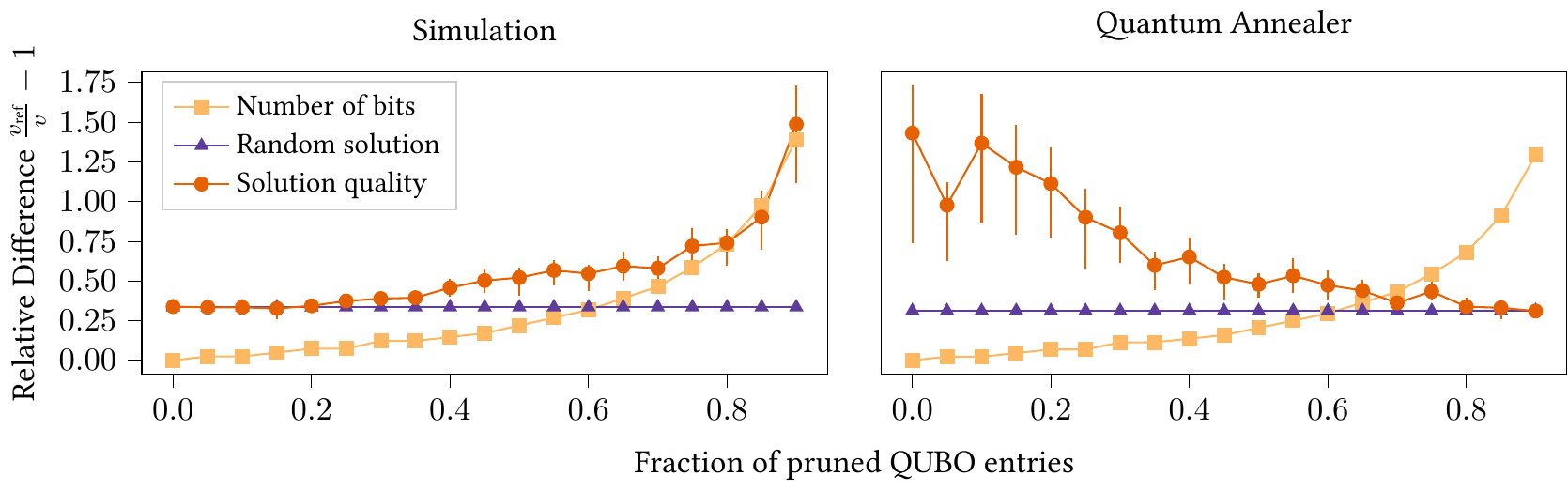}\vspace*{-1em}
    \caption{Ratio of solutions ($v$) compared to optimum ($v_{\textrm{ref}}$) and ratio of embeddable instance size ($v$) compared to the original, non-pruned instance ($v_{\textrm{ref}}$) for the Maximum Cut Problem}
    \label{mc:erg}
\end{figure*}
For the \emph{threshold} strategy, solution quality remains constant up to a pruning factor of 45\% (recall that  all values smaller than 45\% of the maximal value in the \QUBO are deleted
in this case). Beyond this regime, solution quality drops 
considerably.\footnote{For EC, any deleted values are even-valued non-diagonal matrix entries. The maximal value of the \QUBO is 8 for our data. A pruning factor $p = {0.0, 0.05, 0.1, 0.15,
0.2}$  multiplied by 8 results in a threshold less than 2, and no no entry in the \QUBO is
deleted. Pruning factors $p = {0.25, 0.3, 0.35, 0.4, 0.45}$ delete all entries with value 2.
For higher pruning factors entries with value 4 are also deleted, which leaves---for our data
set---with less than 10\% of \QUBO the original entries.}

For strategy \emph{random}, the solution quality decreases almost linearly with the
fraction of pruned \QUBO entries. Consequently, method \emph{fraction}
provides the best results even for high pruning factors. Since we observe the same
behaviour for the other problems considered in the paper, we restrict our discussion
to this pruning strategy in the following.

\subsection{Maximum Cut}
The \emph{Maximum Cut} (MC) problem is defined on undirected graphs
$G = (V, E)$ 
and seeks a partition of $V$ into two disjoint sets $V_1$, $V_2$ such that the cut, that is the number of edges connecting the two sets, is maximal. The decision variant of MC is contained in class NP~\cite{kar72}. Formulating of the MC is straightforward using variable $x_v$ that is set 1
if node $v$ is element of $V_1$, and 0 otherwise:
\(\min \sum_{uv \in E} 2x_u x_v - x_u - x_v\)

Figure~\ref{mc:erg} shows the results of evaluating each pruned \QUBO 100 times on a simulation
and the QA. 
The simulation shows a similar behaviour for the size of embeddable instances and solution quality
as for the EC. Up to a pruning fraction of 35\%, solution quality is almost constant, and up to 70\%
pruning fraction, only a moderate deterioration in quality can be observed, despite
the fact the problem instances comprising up to 50\% more logical qubits are
tractable.

Compared to a random solution the solution of the simulation is never better.
The QA obtains solutions of lower quality even when no pruning has yet
been performed. With higher pruning fractions, solutions \emph{improve} and converge to the solution of a random solution.

\begin{figure}[H]
    \centering
    \includegraphics[scale=1.0]{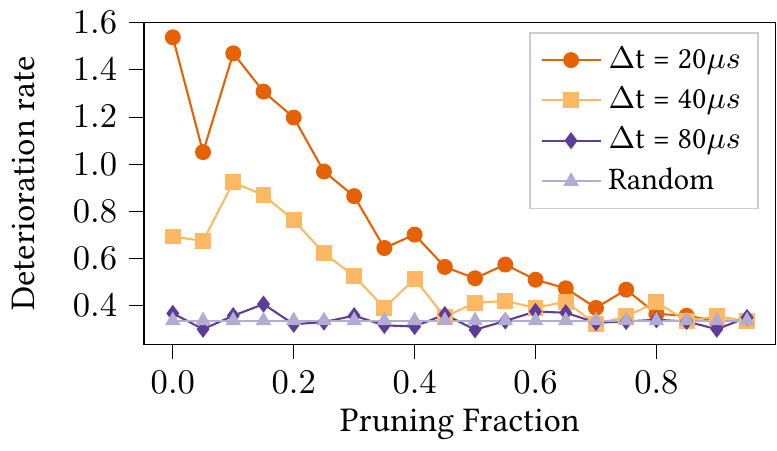}\vspace*{-1em}
    \caption{Solution Quality for Maximum Cut for different annealing times and a random solution compared to optimum}
    \label{mc:pruning}
\end{figure}

\begin{figure*}
		\centering 
		\includegraphics{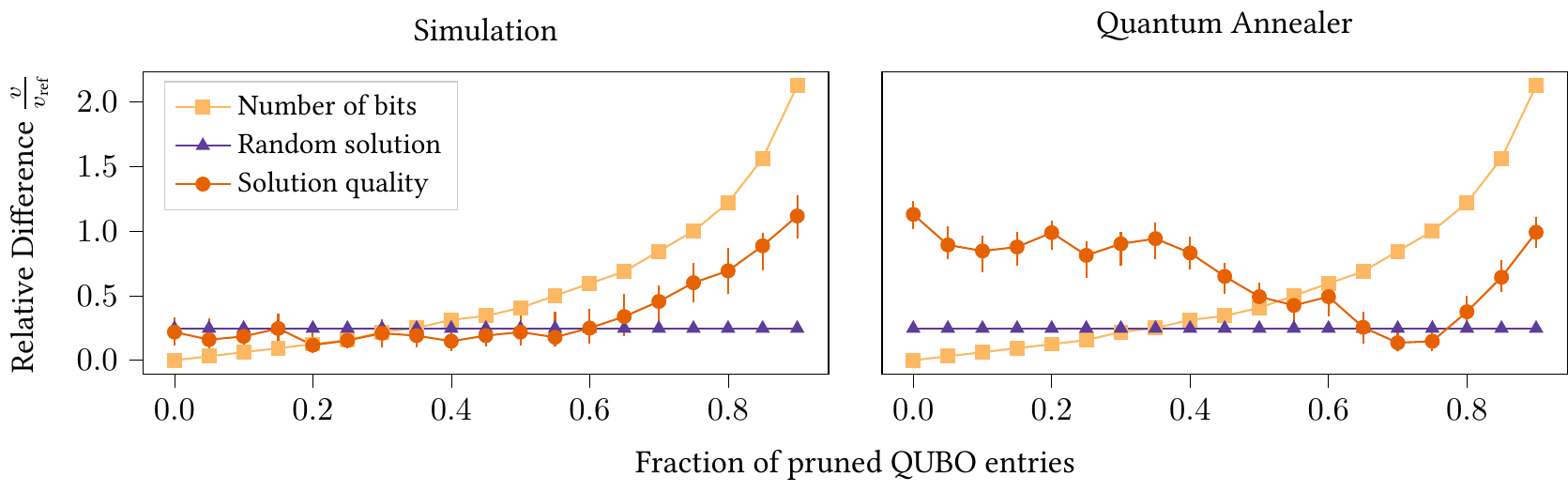}\vspace*{-1em}
        \caption{Ratio of solution ($v$) compared to half sum of elements in set $S$ ($v_{\textrm{ref}}$) and ratio of embeddable instance size ($v$) compared to the original, non-pruned instance ($v_{\textrm{ref}}$) for the Number Partitioning  Problem}
	\label{erg:np}
\end{figure*}
The effect of adjusting physical parameters of the annealer, in particular the annealing
time $\Delta t$, is shown in Figure~\ref{mc:pruning}.
Without any pruning, the standard annealing time of 20$\mu s$ delivers \emph{worse} solutions than
the simulation. Increasing the annealing time to 40$\mu s$ or even 80$\mu s$ provides substantially
better solutions especially for less pruned \QUBOs. With a higher pruning fraction the solutions step-by-step converge to a randomly guessed solution. An optimal value was never reached even with adjusting the annealing time.

\subsection{Number Partitioning}
The \emph{Number Partitioning Problem}, one of Karp's seminal 21 NP-complete problems~\cite{kar72},
asks if a set $S = \{n_1, ..., n_N\}$ of numbers can be divided into two subsets $S_1$, $S_2$
such that $S_1 \cup S_2 = S$, $S_1 \cap S_2 = \emptyset$ and $\sum_{n_i \in S_1} n_i = \sum_{n_i \in S_2} n_i$.
For defining the Number Partitioning problem as an optimisation problem we use the objective
function $\min \sum_{n_i \in S_1} n_i - \sum_{n_i \in S_2} n_i$, that is minimising the difference
between sets $S_1$ and $S_2$.

Following~\cite{luc14}, a \QUBO formulation of the problem is given by
\begin{equation}
\min A\;\; \left( 2\sum_{i=1}^N \sum_{j>i}^N 4x_i n_i x_j n_j + \sum_{i=1}^N 4x_in_i^2 - 4k\sum_{i=1}^N x_in_i + k^2 \right)
\end{equation} 
with $k = (\sum_{i=1}^N)^2$.

The results of approximating this \QUBO by the \emph{fraction} strategy are shown in Figure~\ref{erg:np}.
Since we minimize the difference between the sum of sets $S_1$, $S_2$, an optimal solution has got 
size 0. To define a relative quality measure, we compared the sum $v$ for a given solution 
to half the sum $v_{\textrm{ref}}$ of all numbers in $S$ by dividing $\frac{v}{v_{\textrm{ref}}}$. For the simulation, deterioration of solution quality
and the growing size of the embeddable instances follow the same trend, similar to the
simulation of MC. The Quantum Annealer shows a \emph{different} behaviour: Solutions for
the original, non-pruned \QUBO are \emph{worse} on the QA than the same solution on a
simulation. Higher pruning fractions lead to better solutions on the QA, also exceeding a randomly
guessed solution.

\begin{figure}
    \centering
    \includegraphics[scale=1.0]{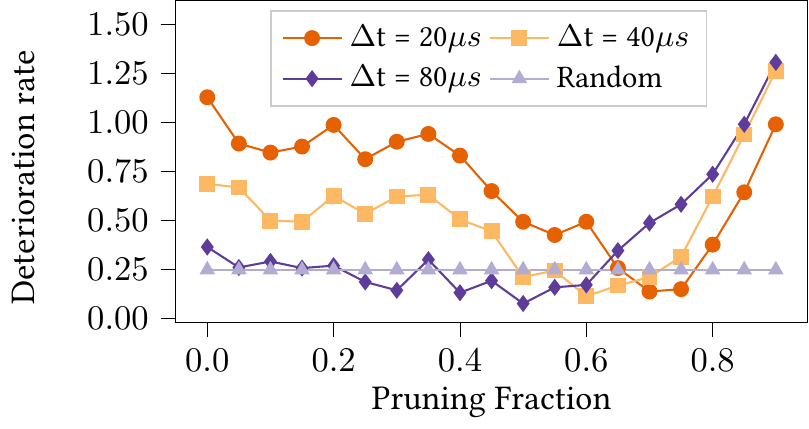}
    \caption{Ratio of solutions to mean sum of numbers in $S$}
    \label{np_pruning_graph}
\end{figure}

Figure~\ref{np_pruning_graph} shows the solution quality of runs with different annealing times compared to the half sum of numbers, that have to be partitioned. 
The solution quality of 100 runs with 20 and 40$\mu s$ are almost identical, thus showing that the adjustment of the annealing time does not change the solution quality for number partitioning. Nevertheless the solutions are mostly worse than the solutions of a simulation. With increasing pruning fraction of more than 80\% the quality of solutions drops significantly for all annealing times. 

\subsection{Airport Gate Assignment Problem}
The \emph{airport gate assignment problem} (AGAP) is a specialisation
of the well-known quadratic assignment problem (QAP). It is contained
in \APX~\cite{sah76}. For a given airport with $m$ gates and $n$
airplanes, the task is to find an assignment between airplanes and
gates so that walk way between gates for passengers changing between
flights is minimal (of course, every airplane is assigned to exactly
one gate, and no two planes are assigned to the same gate). Related to ~\cite{din03} this problem is mathematically described by:

\begin{align}
\min & \sum_{i = 1}^n \sum_{j = 1}^n \sum_{k = 1}^m \sum_{l = 1}^m 
( p_{i,j} d_{k,l} + a_{i, k})  x_{i,k} x_{j,l}  + \nonumber\\
	&\sum_{i = 1}^n \sum_{k = 1}^m p_{0, i} d_{0, k}  x_{i, k} + \sum_{i = 1}^n \sum_{k = 1}^m p_{i, n+1} d_{k, m+1} x_{i, k} + \nonumber \\
	&A \;\sum_{i}^n \left( \sum_{k}^m x_{i,k} - 1 \right)^2 +
	B \; \sum_{k}^m \left( \sum_{i}^n x_{i,k} - 1 \right) ^2  
\label{mini}
\end{align}

\begin{figure*}
\centering
\includegraphics[scale=0.98]{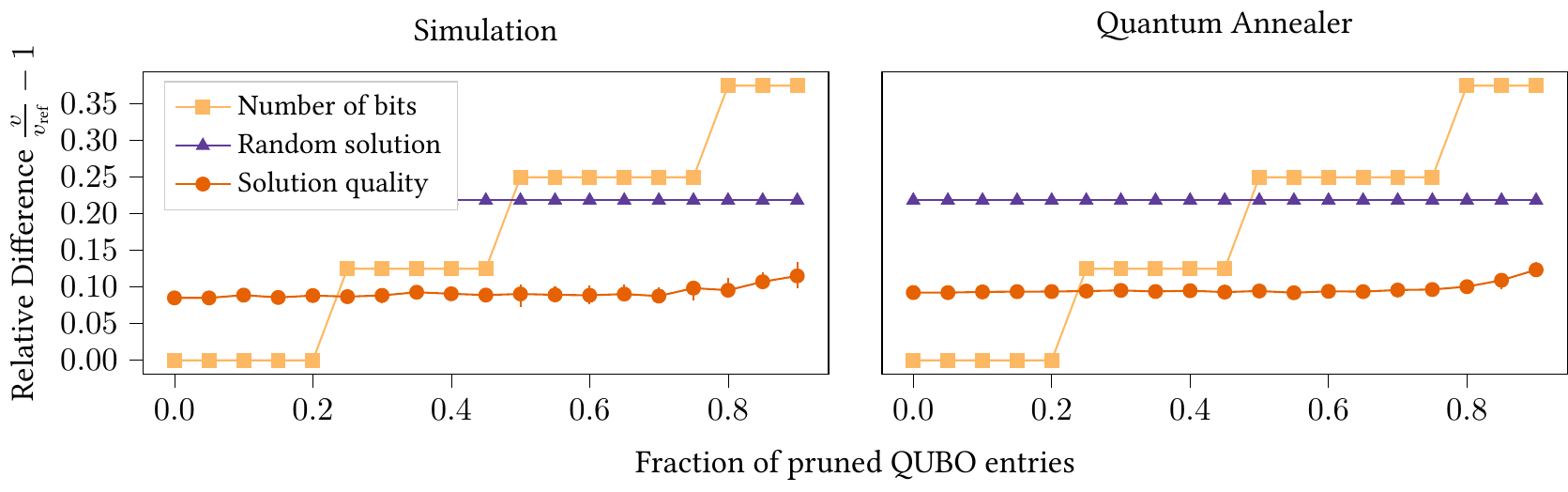}\vspace*{-1em}
\caption{Ratio of solutions ($v$) compared to optimal value ($v_{\textrm{ref}}$) of the instance and ratio of embed-able instance size ($v$) compared to the original, non-pruned instance ($v_{\textrm{ref}}$) for the Airport Gate Assignment Problem. }
\label{agap:pruning}
\end{figure*}

\begin{figure*}
    \centering
    \includegraphics[scale=0.98]{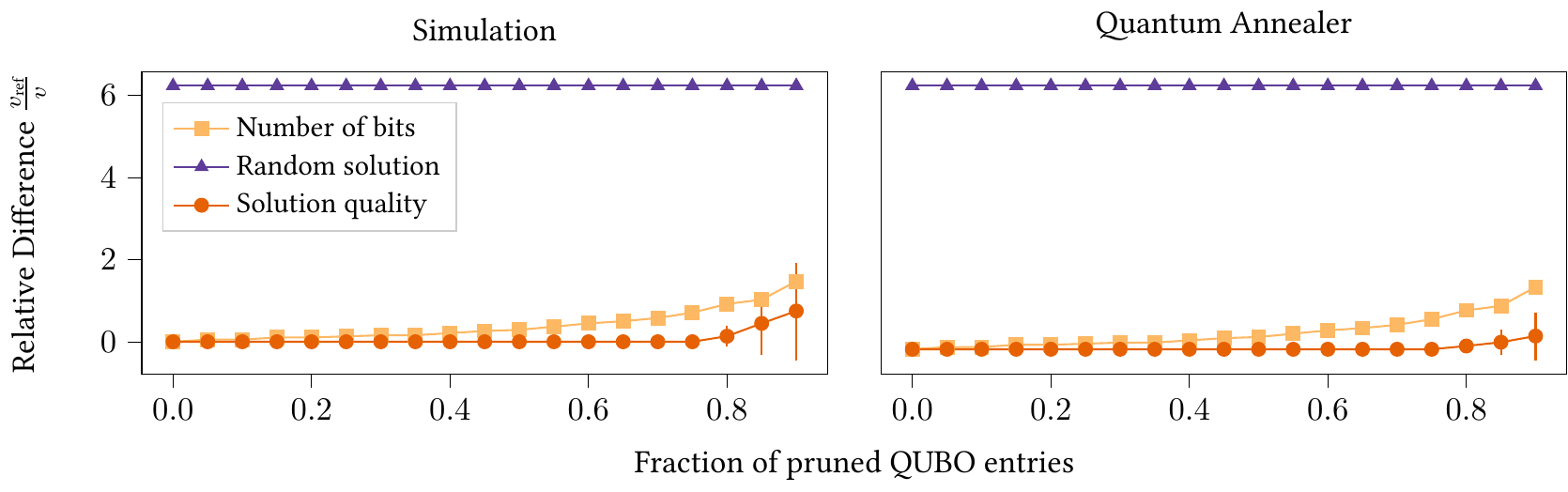}\vspace*{-1em}
    \caption{Ratio of correctly solved clauses ($v$) compared to number of clauses  ($v_{\textrm{ref}}$) and ratio of embeddable instance size ($v$) compared to the original, non-pruned instance ($v_{\textrm{ref}}$) for the Maximum 3-SAT Problem}
    \label{fig:3sat}
\end{figure*}

Binary variable $x_{i,k}$ is set 1 if plane $i$ is assigned to gate
$k$, and 0 otherwise. $p_{i,j}$ describes the number of passengers changing from
plane $i$ to plane $j$ (plane 0 represents a dummy-plane for
passengers boarding their initial flight in a multi-leg trip, and
dummy-plane $n+1$ collects passengers leaving the airport after their
final leg). $d_{k,l}$ specifies the distance between gate $k$ and
$l$. Variable $a_{i,k}$ provides costs for assigning plane $i$ to gate
$k$. Parameter $y_{i,j}$ describes whether two planes can be assigned to the same gate because they will be on the gate at different times.
$A,B \in \mathbb{R}$ are relative weights that ensure that violating a
constraint cannot lead to optimal, but incorrect solutions.

Figure~\ref{agap:pruning} shows the result of approximation by pruning, based on 200 runs for each
reduced \QUBO. 
In Figure~\ref{agap:pruning} we show the deviation of solutions for simulation and
QA calculation compared to the optimal value of the problem instance. Both, QA and
simulation, find better solutions than random guessing, and the solution quality
is almost constant independent of the degree of approximation. When pruning more
than 80\% of the \QUBO matrix, the solution quality drops to a deviation of 15\% from
the optimal value. However, using the strongly pruned \QUBO, it is possible to
embed about 30\% larger problems on the quantum hardware.

Problem AGAP is well suited for approximation; since the underlying 
quadratic assignment problem finds application various industrial relevant contexts,
including placement problems, scheduling applications or parallel and distributed
computing~\cite{cel13}.

\subsection{Maximum 3SAT}
Boolean satisfiability with three literals per clause is the
cornerstone problem of NP completeness~\cite{coo71}, and the
corresponding approximation problem is in the hard approximation class
APX~\cite{pap91}. Given a Boolean formula with $n$ variables and $m$
clauses, where each clause contains a disjunction of exactly three
literals, the problem is to find an assignment of the $n$ variables
that satisfies as many clauses as possible. The decision variant, the 3-SAT Problems, asks whether a given formula can be assigned with $n$ variables such that the whole formula, that is all clauses contained in the formula, is satisfied. Applications of the maximum 3SAT problem are found, for instance, in database systems, combinatorial optimization, or expert systems~\cite{che02}.

\begin{figure*}
    \centering
    \includegraphics[scale=0.98]{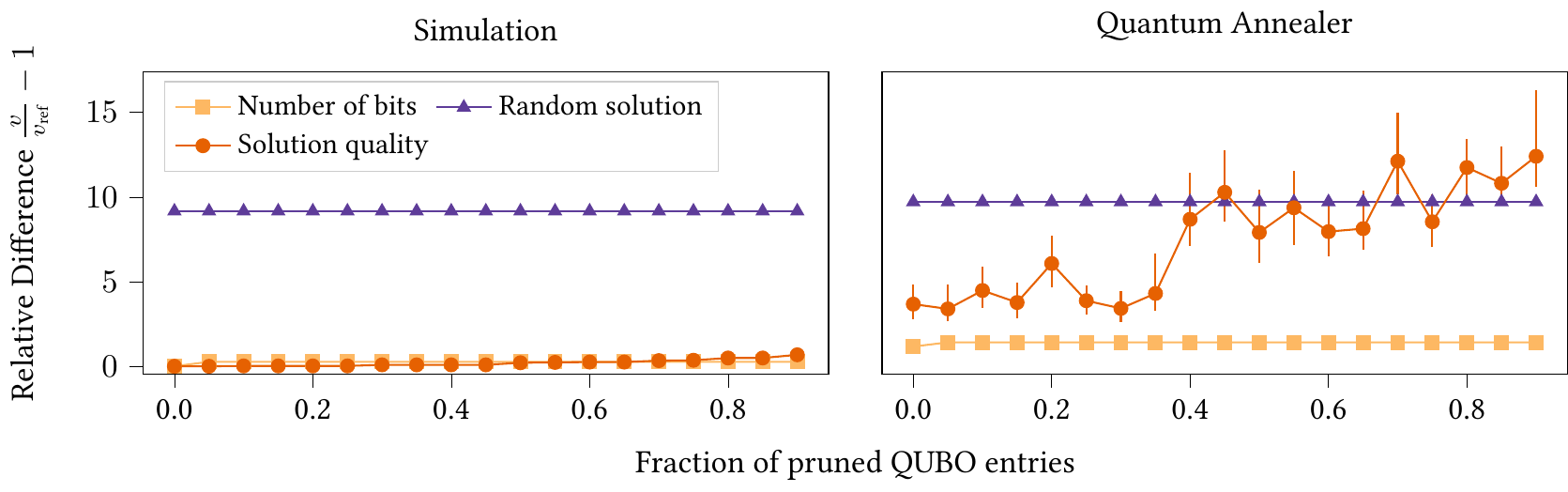}\vspace*{-1em}
    \caption{Ratio of solutions ($v$) compared to optimum ($v_{\textrm{ref}}$) and ratio of embeddable instance size ($v$) 
    compared to the original, non-pruned instance ($v_{\textrm{ref}}$) of the travelling salesperson problem.}
    \label{tsp:pruning}
\end{figure*}

Following~\cite{cho10}, we obtaining a \QUBO for 3SAT via the weighted maximum
independent set (WMIS) problem. It is known that for solving randomly created SAT instances, the ratio $\alpha_c=m/n$ is an
important characteristic quantity that influences the hardness of the problem.
For small $\alpha_c$, it is on average easy to find a satisfying assignment.

The result of approximation by pruning is shown in Figure~\ref{fig:3sat}.
Solution quality is mostly invariant for up to 70\%. Conflict edges receive larger
weights than other edges in the problem graph, and consequently, small \QUBO entries
are removed first, and conflict edges last.
Consequently, the probability of creating a contradiction (where a
variable and its negated form are both assigned the same truth value)
is small in the initial phase of the pruning process, and the
solution quality is stable until entries representing hard constraints
are affected.

\subsection{Travelling Salesperson}
The \emph{travelling salesperson problem} (TSP) is a well-known NP-com\-plete
problem~\cite{dur87} whose optimisation variant is contained in
NPO~\cite{aus99}.  An input graph $G = (V,E)$ with
$N = \abs{V} $ and weight $W_{uv}$ for every edge $(u,v) \in E$ and
a start node $s$ is given. The TSP seeks a tour through
all the nodes, that is, a path $p$ that starts and ends at node $s$,
and contains every other node exactly once. The total weight of the
path given by $ \sum_{(u,v) \in p} W_{uv}$ must be minimised.

Following Lucas~\cite{luc14}, a QUBO for
the problem is given by
\begin{align}
  \min\ &A\; \sum_{v=1}^n (1 -\sum_{j=1}^N x_{v,j})^2+ \nonumber
A\; \sum_{j=1}^n(1-\sum_{v =1}x_{v,j})^2 + \nonumber\\
&A\; \sum_{(uv) \not \in E} \sum_{j=1}^N x_{v,j}x_{v,j+1} + 
\; \sum_{(uv) \in E} W_{uv} \sum_{j=1} x_{v,j}x_{v,j+1}\label{eq:tsp}
\end{align}
We use binary variable $x_{v,j} $ to indicate if node $v$ is visited on the 
$j$-th position in the path. The first two terms in Eq.~(\ref{eq:tsp})
ensure that every node 
is visited exactly once within the tour and that there has to be a $j$-th node in the 
path. The third term prevents that two unconnected nodes appear consecutively in the
path; we assume for our analysis that all nodes are 
connected to each other, and can omit the term. The last term minimises the overall 
weight of the selected path by adding all weights of the edges contained in the path. 
Values $A,B$ are positive integers that must satisfy
$0<B\max(W_{uv}) < A$.

The results of the experiment are given in Figure~\ref{tsp:pruning}. For the TSP, simulation and QA
exhibit completely different behaviour. Up to a pruning fraction of 80\%, the solution quality in the
simulation is almost invariant. For the QA, solution quality drops considerably after a pruning
degree of only 20\%. Most importantly, the QA did not obtain any \emph{valid}
solutions, which explains the strong fluctuations in solution quality.

\subsection{Graph Coloring}
The \emph{Graph Coloring problem} (GC) is another of Karp's 21 NP-complete problems~\cite{kar72}. The decision variant of GC decides if the nodes $V$ of a graph $G = (V, E)$ can be colored in such a way that no edge $e \in E$ connects two nodes of the same color. One common objective for an optimisation
variant is to minimise the number of required colors. 
Ref.~\cite{kud18} solves small instances of GC using a new approach called constrained quantum annealing, which substantially reduces the dimension of the solution space.

For approximate solutions, the number of errors is given by the amount of edges connecting
two same-colored nodes. Following~\cite{luc14}, a \QUBO formulation of GC is given by
\begin{equation}
\min A\; \sum_{v \in V} (1 - \sum_{i=1}^n x_{v,i}) ^ 2 +
    B \; \sum_{uv \in E} \sum_{i=1}^n x_{u,i}x_{v,i}
    \label{eq:gc}
\end{equation}
Binary variable $x_{v_i}$ indicates that node \(v\) has color \(i\);  $n$ gives the number of possible
colors. The first term ensures that every node is assigned exactly one color,
and the second term minimises the number of errors (ideally, zero).

\begin{figure*}
    \centering
    \includegraphics[scale=0.98]{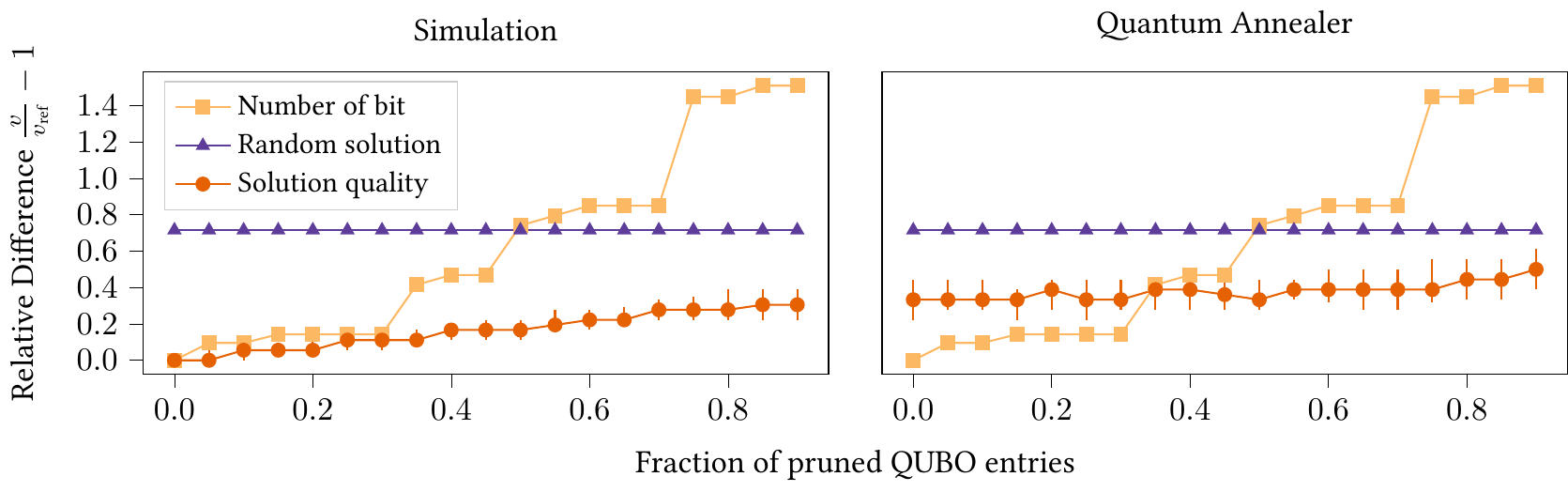}\vspace*{-1em}
    \caption{Ratio of errors ($v$) compared to number of nodes in the graph ($v_{\textrm{ref}}$) and ratio of
    embeddable instance size ($v$) compared to the original, non-pruned instance ($v_{\textrm{ref}}$) for the
    graph coloring problem.}
    \label{gc:pruning}
\end{figure*}

Running the pruned \QUBO 100 times in a simulation and a Quantum annealer gives results as shown in
Figure~\ref{gc:pruning}. We only delete values that represent the second term of the
minimisation formula in~\ref{eq:gc}. The hard constraints that ascertain that assignment of
exactly one color to every node remain.

The simulation shows that approximation leads to only moderate loss in solution quality up to a pruning
fraction of 50\%, which is desirable. The obtained solutions improve over the random choice
baseline. The QA did not find any valid solutions, which  means that one node was either assigned two
colors, or no color at all.

\begin{figure*}[htb]
    \centering
    \includegraphics[scale=0.98]{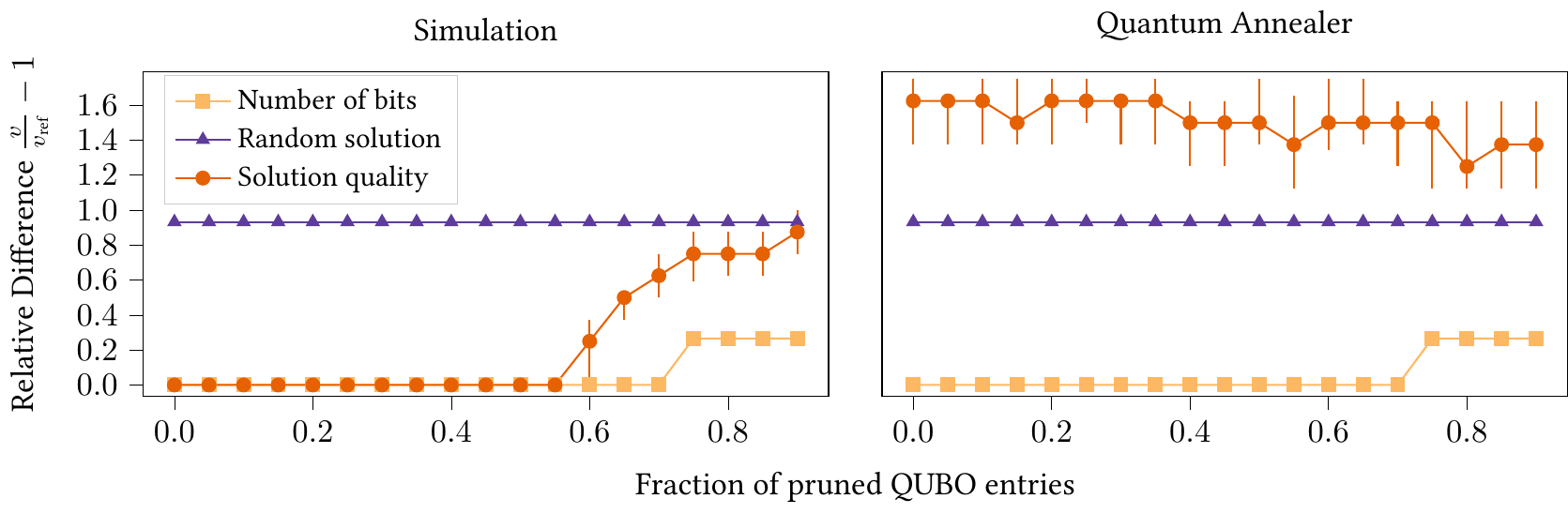}\vspace*{-1em}
    \caption{Ratio of errors ($v$) compared to number of nodes in the graph ($v_{\textrm{ref}}$) and ratio of embeddable
    instance size ($v$) compared to the original, non-pruned instance ($v_{\textrm{ref}}$) for the graph isomorphism problem.}
    \label{gi:pruning}
\end{figure*}

\subsection{Graph Isomorphism}
Given two graphs $G_1 = (V_1, E_1)$ and $G_2 = (V_2, E_2)$, the \emph{graph isomorphism problem}
(GI) decides if there exists a permutation $P$ such that the adjacency matrices $A_1$, $A_2$ for
graphs $G_1$, $G_2$ are related by $A_2 = P^TA_1P$~\cite{luc14}.  
We consider the optimisation variant of the problem, that maximises how many vertices of
$G_1$ can be mapped onto vertices of $G_2$. 

Lucas~\cite{luc14} provides the \QUBO formulation
\begin{align}
  \min\ &A\; \sum_{v} (1 -\sum_{i} x_{v,i})^2 + A\; \sum_{i}(1-\sum_{v }x_{v,i})^2 + \nonumber\\
&B\; \sum_{(ij) \not \in E_1} \sum_{(uv) \in E_2} x_{u,i}x_{v,j} + 
B\; \sum_{(uv) \in E_1} \sum_{(uv) \not \in E_2} x_{u,i}x_{v,j}\label{eq:gi}
\end{align}
Binary variable $x_{v,i}$ is set 1 if node $v$ from $G_1$ is mapped to node $i$ from $G_2$. The
first two terms ensure that every node from $G_1$ is assigned to exactly one node from $G_2$
and vice versa. The last two terms describe how many edges from one graph cannot be found
in the other graph and must be minimised.
As the hard constraints consume a substantial portion of the \QUBO, the size of the embedding does not decrease significantly with pruning.

The results of 100 simulation and QA runs per pruned \QUBO are shown in Figure~\ref{gi:pruning}.
Similar like for the GC problem, the solutions of simulation and QA clearly differ. While 
solution quality stays invariant to the pruning of 50\% in the simulation, QA could not
find any valid solutions at all. Additionally, the growth in size of embeddable instances with
increasing pruning fraction is negligible, which makes GI not well suited for approximation.

\section{Conclusion}
We have introduced several approximation methods for optimisation problems in QUBO formulation
that allow us to trade decreasing solution quality for a smaller amount of
required qubits. We have compared the behaviour of  solution qualities for problems from different
approximation complexity classes using a classical simulation and a quantum annealer.

We find that the achievable solution quality on QA is robust against pruning \QUBO matrices,
often up to pruning ratios as large as 50\% or more. Since QAs are probabilistic machines by
design, they usually deliver sub-optimal, approximate results anyway, the loss in solution
quality is only of subordinate relevance, and is compensated by the fact that pruned \QUBO
matrices allow for handling larger problem instances on hardware of a given capacity.

We also observe that for many of the problems analysed in this paper, QA without postprocessing
delivers solutions that are either close to, or even below the quality of randomly guessed
solutions, effectively eliminating any quantum-mechanical advantage. However, as the results
obtained by classical simulations show, approximate solutions can---given suitable future hardware---deliver solutions of comparable quality to the full problem description, while
remarkably reducing the amount of required qubits. Since the amount of qubits will remain one
major limiting factor on real hardware, our results might be useful to enlarge the possible
problem sizes treatable on such machines, hopefully assisting first real-world industrial 
applications of quantum computing.

%
%
%
 \clearpage\bibliographystyle{splncs04}
 \bibliography{literatur}

\begin{thebibliography}{10}
\providecommand{\url}[1]{\texttt{#1}}
\providecommand{\urlprefix}{URL }
\providecommand{\doi}[1]{https://doi.org/#1}

\bibitem{aar13}
Aaronson, S.: Quantum Computing since Democritus. Cambridge University Press
  (2013)

\bibitem{aru19}
Arute, F., Arya, K., Babbush, R., et~al: Quantum supremacy using a programmable
  superconducting processor. Nature 574, 505–510  (2019)

\bibitem{aus99}
Ausiello, G., Crescenzi, P., Gambosi, G., Kann, V., Marchetti-Spaccamela, A.,
  Protasi, M.: Complexity and Approximation - combinatorial Optimization
  Problems and their Approximability Properties. Springer-Verlag Berlin (1999)

\bibitem{ren19}
van Bevern, R., Tsidulko, O.Y., Zschoche, P.: Fixed-parameter algorithms for
  maximum-profit facility location under matroid constraints. In: Heggernes, P.
  (ed.) Algorithms and Complexity. Springer International Publishing (2019)

\bibitem{cas17}
Catelvecchi, D.: Quantum cloud goes commercial. Nature  (2017)

\bibitem{cel13}
Cela, E.: The Quadratic Assignment Problem: Theory and Algorithms. Springer
  Science and Business Media (2013)

\bibitem{che02}
Chen, J., Kanj, I.A.: Improved exact algorithms for max-sat. In: LATIN 2002:
  Theoretical Informatics. Springer Berlin Heidelberg (2002)

\bibitem{cho10}
Choi, V.: Adiabatic quantum algorithms for the {NP}-complete maximum weight
  independent set, exact cover and {3SAT} problems. arXiv:1004.2226  (2010)

\bibitem{coo71}
Cook, S.A.: The complexity of theorem-proving procedures. Proceedings of the
  third annual ACM symposium on Theory of computing  (1978)

\bibitem{cyg19}
Cygan, M., Lokshtanov, D., Pilipczuk, M., Pilipczuk, M., Saket, S.: Minimum
  bisection is fixed-parameter tractable. SIAM J. Comput., 48(2)  (2019)

\bibitem{dwa18}
{D-Wave Systems Inc.}: Getting started with the {D-Wave} - {U}ser {M}anual

\bibitem{dev19}
Devanur, N.R., Jain, K., Sivan, B., Wilkens, C.A.: Near optimal online
  algorithms and fast approximation algorithms for resource allocation
  problems. J. ACM  (Jan 2019)

\bibitem{din03}
Ding, H., Lim, A., Rodrigues, B., Zhu, Y.: The airport gate assignment problem.
  Decision Science Institute Annual Meeting, Washington, DC  (2003)

\bibitem{din19}
Ding, Y., Chen, X., Lamata, L., Solano, E., Sanz, M.: Logistic network design
  with a d-wave quantum annealer (2019)

\bibitem{dur87}
Durbin, R., Willshaw, D.: An analogue approach to the travelling salesman
  problem using an elastic net method. Nature, 326(6114):689  (1987)

\bibitem{fel11}
Feldman, M.: D-wave sells first quantum computer. HPCwire  (2011)

\bibitem{glo18}
Glover, F.W., Kochenberger, G.A.: A tutorial on formulating qubo models. ArXiv
  \textbf{abs/1811.11538} (2018)

\bibitem{gyo19}
Gyongyosi, L., Imre, S.: A survey on quantum computing technology. Computer
  Science Review(37)  (2019)

\bibitem{kar72}
Karp, R.: Reducibility among combinatorial problems. Complexity of Computer
  Computations  (1972)

\bibitem{kud18}
Kudo, K.: Constrained quantum annealing of graph coloring. Phys. Rev. A
  \textbf{98} (2018)

\bibitem{lew17}
Lewis, M.W., Glover, F.: Quadratic unconstrained binary optimization problem
  preprocessing: Theory and empirical analysis. CoRR  (2017)

\bibitem{luc14}
Lucas, A.: Ising formulation of many {NP}-problems. Frontiers in Physics
  (2014)

\bibitem{mcg14}
McGeoch, C.: Adiabatic quantum computation and quantum annealing: Theory and
  practice. Synthesis Lectures on Quantum Computing  (2014)

\bibitem{nie00}
Nielson, M.A., Chuang, I.L.: Quantum Computation and Quantum Information.
  Cambridge University Press (2000)

\bibitem{pap91}
Papadimitriou, C.H., Yannakakis, M.: Optimization, approximation, and
  complexity classes. J. Computer and System Sciences 43  (1991)

\bibitem{sah76}
Sahni, S.K., Gonzalez, T.F.: P-complete approximation problems. J. ACM 23
  (1976)

\bibitem{sto19}
{Stollenwerk}, T., {O'Gorman}, B., {Venturelli}, D., {Mandrà}, S.,
  {Rodionova}, O., {Ng}, H., {Sridhar}, B., {Rieffel}, E.G., {Biswas}, R.:
  Quantum annealing applied to de-conflicting optimal trajectories for air
  traffic management. IEEE Transactions on Intelligent Transportation Systems
  (2019)

\bibitem{str19}
Streif, M., Neukart, F., Leib, M.: Solving quantum chemistry problems with a
  d-wave quantum annealer. In: Feld, S., Linnhoff-Popien, C. (eds.) Quantum
  Technology and Optimization Problems. Springer International Publishing
  (2019)

\bibitem{ven15}
Venturelli, D., Mandrà, S., Knysh, S., O’Gorman, B., Biswas, R.,
  Smelyanskiy, V.: Quantum optimization of fully connected spin glasses.
  Physical Review  (2015)

\bibitem{wil19}
Wilson, M.L., Vandal, T., Hogg, T., Rieffel, E.G.: Quantum-assisted associative
  adversarial network: Applying quantum annealing in deep learning. ArXiv
  \textbf{abs/1904.10573} (2019)

\end{thebibliography}

\end{document}